\documentclass[a4paper,accepted=2020-02-24]{quantumarticle}
\pdfoutput=1
\usepackage[numbers,sort&compress]{natbib}

\usepackage{amsmath}
\usepackage{amssymb}
\usepackage{graphicx}
\usepackage{bm}
\usepackage{braket}
\usepackage{bbm}

\begin{document}

\title{Engineering Schr\"{o}dinger cat states with a photonic even-parity detector}

\author{G. S. Thekkadath}
\affiliation{Clarendon Laboratory, University of Oxford, Parks Road, Oxford, OX1 3PU, UK}

\author{B. A. Bell}
\affiliation{Clarendon Laboratory, University of Oxford, Parks Road, Oxford, OX1 3PU, UK}

\author{I. A. Walmsley}
\affiliation{Clarendon Laboratory, University of Oxford, Parks Road, Oxford, OX1 3PU, UK}

\author{A. I. Lvovsky}
\affiliation{Clarendon Laboratory, University of Oxford, Parks Road, Oxford, OX1 3PU, UK}
\affiliation{Russian Quantum Center, 100 Novaya St., Skolkovo, Moscow 143025, Russia}

\begin{abstract}
When two equal photon-number states are combined on a balanced beam splitter, both output ports of the beam splitter contain only even numbers of photons.
Consider the time-reversal of this interference phenomenon: the probability that a pair of photon-number-resolving detectors at the output ports of a beam splitter both detect the same number of photons depends on the overlap between the input state of the beam splitter and a state containing only even photon numbers.
Here, we propose using this even-parity detection to engineer quantum states containing only even photon-number terms. 
As an example, we demonstrate the ability to prepare superpositions of two coherent states with opposite amplitudes, i.e. two-component Schr\"{o}dinger cat states.
Our scheme can prepare cat states of arbitrary size with nearly perfect fidelity. 
Moreover, we investigate engineering more complex even-parity states such as four-component cat states by iteratively applying our even-parity detector.
\end{abstract}

\maketitle

\section{Introduction}
The number of excitations in an optical field determines a fundamental property known as parity.
If a field has an even (odd) number of photons, it is said to have even (odd) parity.
With the exception of the vacuum state, fields in classical (e.g. coherent, thermal) states possess uncertainty in their parity, i.e. they have a non-zero probability to have both even and odd photon numbers.
Conversely, states of light with a definite parity have non-classical features.
For example, squeezed vacuum is a superposition of only even photon numbers and has reduced quantum fluctuations in its electric field compared to classical light~\cite{lvovsky2015squeezed}.
This reduction in noise makes squeezed vacuum a valuable resource for optical quantum information processing~\cite{weedbrook2012gaussian,braunstein2005quantum} and quantum sensing~\cite{caves1981quantum}.
Other notable examples of definite parity states that have found uses in quantum technologies include Schr\"{o}dinger cats~\cite{schrodinger1935gegenwartige}, Holland-Burnett~\cite{holland1993interferometric}, and Gottesman-Kitaev-Preskill states~\cite{gottesman2001encoding}.
The ability to prepare these and other definite parity optical states in a scalable and robust manner is highly desirable for developing quantum technologies.

On-demand manipulation of the parity of an optical field is not an easy task since it requires non-linearities at the single-photon level.
One way to achieve these non-linearities is to couple the field to atoms~\cite{yurke1986generating,law1996arbitrary,gerry2005quantum,hacker2019deterministic}.
An alternative approach is to perform a measurement (e.g. photon number, quadrature) on one subsystem of an entangled state to conditionally prepare the state of the other subsystem.
Because this approach is easier in practice, there have been many proposals and experimental demonstrations of measurement-based state engineering~\cite{lvovsky2002catalysis,zavatta2004quantum,fiurasec2005conditional,bimbard2010quantum,sperling2014quantum}, including those focusing on Schr\"{o}dinger cat states~\cite{glancy2008methods,dakna1997generating,neergaard2006generation,ourjoumtsev2006generating,ourjoumtsev2007generation,wakui2007photon,marek2008generation,takahashi2008generation,huang2015optical,sychev2017enlargement, etesse2015experimental}. 
However, the measurement in most of these schemes is performed by binary ``click" detectors which cannot distinguish between one and many photons.

Thanks to recent advances in detector technology, it is now possible to count the number of photons in an optical field using photon-number-resolving detectors~\cite{lita2008counting}.
This advancement enables new state engineering schemes that exploit the number resolution of such detectors~\cite{karimi2018superposition,quesada2019simulating,su2019generation,su2019conversion}.
For example, some recent proposals and experiments have employed multi-photon subtraction~\cite{gerrits2010generation}, addition~\cite{mccusker2009efficient}, and catalysis~\cite{bartley2012multiphoton,birrittella2018photon,eaton2019gkp}.

In this paper, we propose a novel scheme to engineer a broad class of even-parity states using photon-number-resolving detectors.
We first construct a device that measures the overlap of its input with an even-parity state whose photon-number coefficients are controlled by an ancilla state.
We refer to this measurement device as an ``even-parity detector".
Then, by measuring one of the modes of a two-mode squeezed vacuum state with our even-parity detector, we produce an even-parity state $\ket{\psi}$ in the second mode.
We show that for an ancilla in a coherent state of amplitude $\beta$, the produced even-parity state is a symmetrized version of the ancilla, i.e. $\ket{{\rm cat}_\beta} = \mathcal{N}\left(\ket{\beta} + \ket{-\beta}\right)$, where the two phase-conjugated coherent states, $\ket{\beta}$ and $\ket{-\beta}$, play the role of the ``alive" and ``dead" cats in Schr\"odinger's famous Gedankenexperiment~\cite{schrodinger1935gegenwartige}. 
Cat states have been extensively studied in quantum physics due to their foundational importance~\cite{schleich1991nonclassical,sanders1992entangled,wenger2003maximal,jeong2003quantum,stobinska2007violation,haroche2013nobel,wineland2013nobel,vlastakis2015characterizing,wang2016schrodinger} and their applications in quantum information processing~\cite{enk2001entangled,jeong2001quantum,ralph2003quantum,gilchrist2004schrodinger,lund2008fault,vlastakis2013deterministically,ofek2016extending}.
In principle, our procedure can prepare arbitrarily large cat states with nearly perfect fidelity.
Furthermore, we also investigate engineering four-component cat states which have applications in quantum error correction~\cite{vlastakis2013deterministically,leghtas2013hardware,mirrahimi2014dynamically,ofek2016extending,albert2018performance} and quantum sensing~\cite{zurek2001sub,agarwal2004mesoscopic,dalvit2006quantum,roy2009subplanck}.

\section{An even-parity detector}
\label{sec:even_parity_det}
\begin{figure}
    \centering
    \includegraphics[width=1\columnwidth]{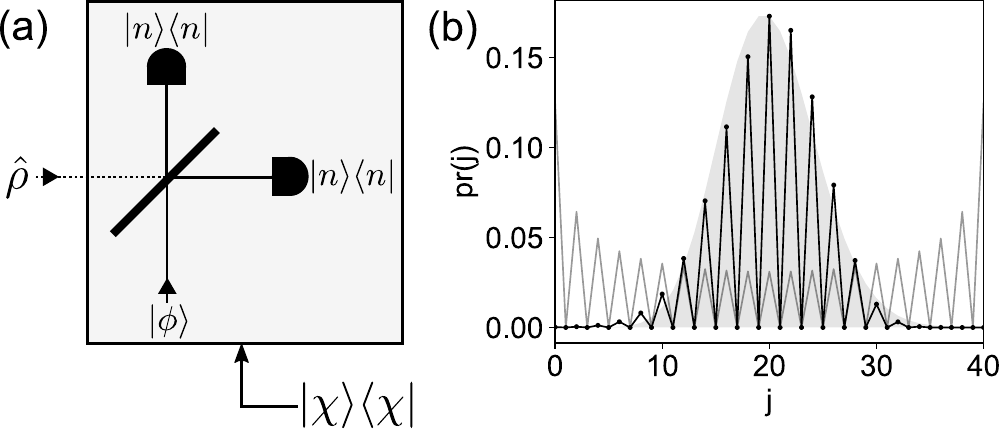}
	\caption{
	The even-parity detector.
	(a) Conditioned on obtaining the detection outcome $(n,n)$, the composite detector shown in the grey box performs the projective measurement $\ket{\chi}\bra{\chi}$ on the input state $\hat{\rho}$. 
	The state $\ket{\chi}$ contains only even photon-number terms whose amplitudes are controlled by $\ket{\phi}$. (b) Example case of $n=20$ and $\ket{\phi}=\ket{\beta}$, a coherent state with amplitude $\beta = \sqrt{20}$. The photon-number distribution $\left|\braket{j|\chi}\right|^2$ (black line) approximates that of the ancilla input $\left|\braket{j|\phi}\right|^2$ (grey area) with the odd photon-number terms eliminated. The grey line shows the matrix element $\left|A_{j,n}\right|^2$ (see Eq.~\eqref{eqn:hb_coeff}) connecting the beam splitter input to the detected state $\ket{n,n}$ at its output.
	Note that the distributions are discrete and the lines are merely to guide the eye.
	}
	\label{fig:Scheme}

\end{figure}

We begin by describing the even-parity detector which is shown in the grey box of Fig.~\ref{fig:Scheme}(a). 
An ``input" state $\hat{\rho}$ is combined with an ancillary ``control" state $\ket{\phi}$ on a balanced beam splitter. 
We assume that $\hat{\rho}$ is arbitrary. 
The control state is a general pure state, which can be written in the photon-number basis as
\begin{equation}
    \label{eqn:control_state}
    \ket{\phi} = \sum_{m=0}^{\infty} c_m \ket{m}
\end{equation}
with $\sum_m |c_m|^2 = 1$.
The outputs of the beam splitter are then sent to photon-number-resolving detectors which we assume to have perfect efficiency for now. 
The joint probability to measure $n$ photons in both output ports, i.e. the outcome $(n,n)$, is given by
\begin{equation}
    \mathrm{pr}(n,n) = \left\langle n,n\left|
    \hat{U}\left[ \hat{\rho}\otimes\ket{\phi}\bra{\phi} \right]\hat{U}^\dagger\right|n,n\right\rangle
    \label{eqn:Pnn}
\end{equation}
where $\hat{U}$ is the balanced beam splitter unitary operator.
Re-writing Eq.~\eqref{eqn:Pnn} as $\mathrm{pr}(n,n) = \braket{\chi|\hat{\rho}|\chi}$, it becomes clear that the measurement device is described by a projector $\ket{\chi}\bra{\chi}$ acting on the input state $\hat{\rho}$ when the detection outcome is $(n,n)$.
The unnormalized projected state $\ket{\chi}$ is given by
\begin{equation}
\ket{\chi} = \braket{\phi|\hat{U}^\dagger|n,n} =\sum_{j=0}^{2n}  c^*_{2n-j} A_{j,n} \ket{j},
\label{eqn:chi_full}
\end{equation}
where
\begin{equation}
\begin{split}
  A_{j,n} &= \braket{j,2n-j|\hat{U}|n,n}\\
  &= \begin{cases} 
  \left(\frac{i}{2}\right)^n  \frac{\sqrt{(2n-j)!(j)!}}{(j/2)!(n-j/2)!} &\textrm{ for even $j$}\\
  0 &\textrm{ for odd $j$}
  \end{cases}
   \label{eqn:hb_coeff}
\end{split}
 \end{equation}
is the matrix element of the beam splitter operator \cite{leonhardt1997measuring}.
As one might expect, $\ket{\chi}$ depends on the photon-number coefficients $c_{2n-j}$ of the control state.
Perhaps more surprisingly, $\ket{\chi}$ consists only of even photon-numbers.
This effect can be understood by considering our device in reverse.
When $\ket{n,n}$ impinges on a beam splitter, a pairing interference effect causes both output ports to only contain even-photon numbers, much like in Hong-Ou-Mandel interference~\cite{hong1987measurement}.
This even-parity state $\hat{U}\ket{n,n}$ was first discussed in Ref.~\cite{campos1989quantum} but is commonly referred to as the Holland-Burnett state after the work of Ref.~\cite{holland1993interferometric}.
By post-selecting the detection outcome $(n,n)$ at the beam splitter output, our even-parity detector destructively projects the two-mode input of the beam splitter onto the Holland-Burnett state with the decomposition into the photon-number basis given by Eq.~\eqref{eqn:hb_coeff}.

In our case, one of the inputs of the beam splitter is the control state $\ket{\phi}$.
As a result, the other input, $\hat{\rho}$, is projected onto an even-parity state $\ket{\chi}$ whose photon-number coefficients are determined by $c_{2n-j}$ of the control state as well as $A_{j,n}$ of the Holland-Burnett state.

Importantly, the latter coefficients are largely constant for $j$ values near $n$.
This can be seen by applying Stirling's approximation to Eq.~\eqref{eqn:hb_coeff}:
\begin{equation}
\begin{split}
    A_{j,n} &\approx \frac{i^n}{\sqrt{\pi}}  \frac{1}{[(j/2)(n-j/2)]^{1/4}} \\
    &\approx \sqrt{\frac{2}{n\pi}}i^n \left(1 + \frac{1}{4n^2}\left(j-n\right)^2 \right) + \mathcal{O}(j-n)^4,
\end{split}
\end{equation}
where in the second line we Taylor expanded $A_{j,n}$ to second order around $j=n$.
In other words, if $j-n\sim\sqrt n$ (which is relevant to the practical case discussed below), the relative variation of $A_{j,n} $ is on a scale of $1/4n$.

In Fig.~\ref{fig:Scheme}(b), we plot the photon-number distribution of the detected state $\ket{\chi}$ when the control state is a coherent state $\ket{\beta}$ with $|\beta|^2=n=20$. This example is of particular interest since $\ket{\chi}$ is approximately the symmetrized version of the control state, i.e. a Schr\"{o}dinger cat state.
This symmetrization occurs for two reasons. Firstly, the photon-number distribution of $\ket{\beta}$ is centered and localized on the flat portion of the photon-number distribution of the Holland-Burnett state, as shown in Fig.~\ref{fig:Scheme}(b).
Secondly, the photon-number distribution of $\ket{\beta}$ is approximately symmetric about the detection outcome $n=20$, i.e. $c_{j}\approx c_{2n-j}$.
As a result, the state $\ket{\chi}$ is given by eliminating the odd photon-number terms of $\ket{\beta}$ while leaving the even terms approximately unchanged.
This operation results in $\ket{\chi} \approx \ket{\beta} + \ket{-\beta}$ since the even (odd) photon-number terms in $\ket{\beta}$ and $\ket{-\beta}$ have equal (opposite) signs.
We discuss the preparation of cat states using our even-parity detector in more detail in Sec.~\ref{sec:cats}.

\subsection{Detection efficiency}
\label{sec:detEff}

\begin{figure}
    \centering
    \includegraphics[width=1\columnwidth]{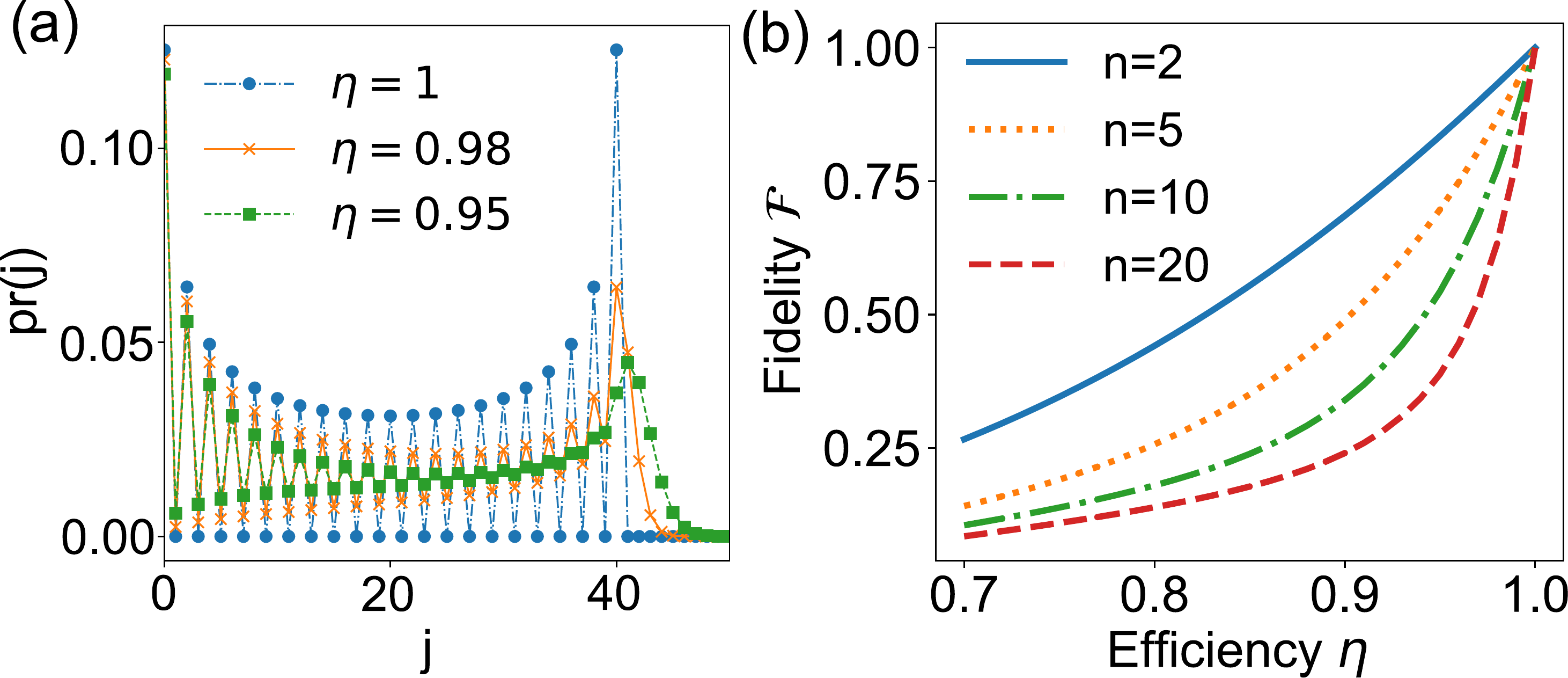}
	\caption{
	Effect of imperfect detection efficiency. (a) The photon-number distribution $\mathrm{pr}(j)$ of the even-parity detector when the detectors have an efficiency $\eta$, i.e. $\mathrm{pr}(j) = \braket{j|\hat{\Pi}(\eta)|j}$. We assume $n=20$ and $c_m \equiv 1$. 
	For $\eta < 1$, odd photon-numbers contribute to $\mathrm{pr}(j)$.
	(b) The fidelity $\mathcal{F} =\braket{\chi|\hat{\Pi}(\eta)|\chi}$ as a function of detection efficiency $\eta$ for various values of $n$.
	}
	\label{fig:detEff}

\end{figure}

We now consider the effects of imperfect detection efficiency on our scheme.
Suppose both photon-number-resolving detectors in Fig.~\ref{fig:Scheme}(a) have an efficiency $\eta$.
Given the outcome $(n,n)$, the even-parity detector projects the input state $\hat{\rho}$ no longer onto a single state $\ket{\chi}$, but rather onto a statistical mixture of states. This measurement is described by an element of a positive operator-valued measure (POVM). 
We derive an expression for this element, $\hat{\Pi}(\eta)$, in Appendix~\ref{app:imp_det_eff}.

In Fig.~\ref{fig:detEff}(a), we plot the photon-number distribution of the imperfect even-parity detector, $\mathrm{pr}(j) = \braket{j|\hat{\Pi}(\eta)|j}$ when $n=20$ and $c_m = 1$ for all $m$, i.e. a control state with a flat photon-number distribution.
Odd photon-number terms quickly begin to contribute to $\mathrm{pr}(j)$ for $\eta < 1$.
To further quantify the effect of loss, we numerically compute\footnote{For the numerical computation, we used the Python packages \emph{Scipy} and \emph{Qutip}~\cite{johansson2013qutip}. 
We generally truncated the Hilbert space to $N=100$.}
the fidelity between $\hat{\Pi}(\eta)$ and the desired projector $\ket{\chi}\bra{\chi}$ using $\mathcal{F} = \braket{\chi|\hat{\Pi}(\eta)|\chi} $ [Fig.~\ref{fig:detEff}(b)].

We see that $\mathcal{F}$ depends strongly on detection efficiency, however less so for smaller $n$ values. This is expected since the probability of the detectors having under-counted at least one photon scales as $(1-\eta)n$, and hence the effects of imperfect efficiency begin to kick in for $(1-\eta)n \gtrsim 1$.
As a point of reference, state-of-the-art transition edge sensor detectors can detect up to $\sim$ 20 photons with $>$ 95\% efficiency~\cite{humphreys2015tomography}.

\section{Even-parity state engineering}

\begin{figure}
    \centering
    \includegraphics[width=0.6\columnwidth]{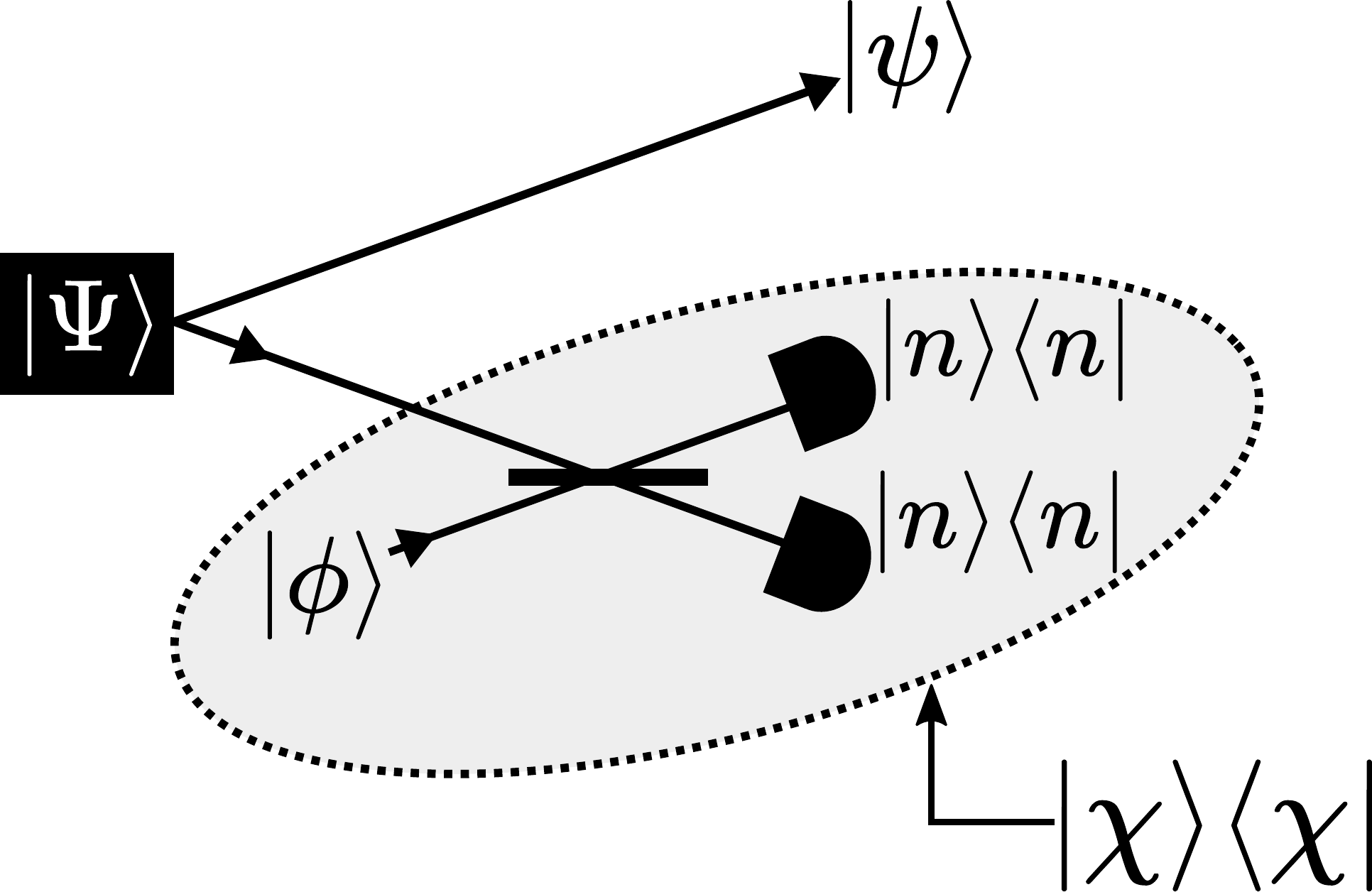}
	\caption{
	Even-parity state engineering. The even-parity detector (shown in grey circle) projects one of the modes of a two-mode squeezed vacuum state $\ket{\Psi}$ onto the state $\ket{\psi} = \braket{\chi|\Psi}$. Just like $\ket{\chi}$, $\ket{\psi}$ is an even-parity state whose photon-number amplitudes are controlled by $\ket{\phi}$. This scheme can be viewed either as remote state preparation of the state $\ket{\chi}$ or as partial teleportation of the state $\ket\phi$.
	}
	\label{fig:stateEng}

\end{figure}

In the form discussed so far, the even-parity detector cannot be used for state engineering since the detected state $\ket{\chi}$ is destroyed upon detection.
Consider instead the scheme shown in Fig.~\ref{fig:stateEng}.
The idea is to use an entangled resource with photon-number correlations to remotely prepare a (possibly imperfect) copy of the state $\ket{\chi}$ in a separate mode. 
We begin with a two-mode squeezed vacuum state,
\begin{equation}
    \ket{\Psi} = \sqrt{1-\lambda^2}\sum_{k=0}^{\infty}\lambda^j \ket{k,k},
\end{equation}
where $\lambda = \tanh{(r)}$ determines the squeezing parameter $r$, which we assume to be positive-real without loss of generality.
Such states can be prepared using various non-linear optical processes such as spontaneous parametric down-conversion using continuous-wave or pulsed pump lasers~\cite{lvovsky2015squeezed}.

By sending one of the modes of $\ket{\Psi}$ to the even-parity detector, the second mode is projected onto the unnormalized state
\begin{equation}
    \ket{\psi} = \braket{\chi|\Psi}=\sqrt{1-\lambda^2}\sum_{j=0}^{2n} c_{2n-j}\lambda^{j} A^*_{j,n}\ket{j}.
    \label{eqn:heralded_state}
\end{equation}

Note that $\ket{\psi}$ is the same as $\ket{\chi}$ except for the factor of $\lambda^{j}$ inside the summation due to the finite squeezing, i.e. $\lambda \neq 1$.
In some cases, it is possible to compensate this effect of finite squeezing when $\lambda$ is known by changing the control state coefficients $\{c_{m}\}$ appropriately.
In the next section, we use this idea to prepare cat states.

\subsection{Two-component cat states}
\label{sec:cats}
\begin{figure}
    \centering
    \includegraphics[width=1\columnwidth]{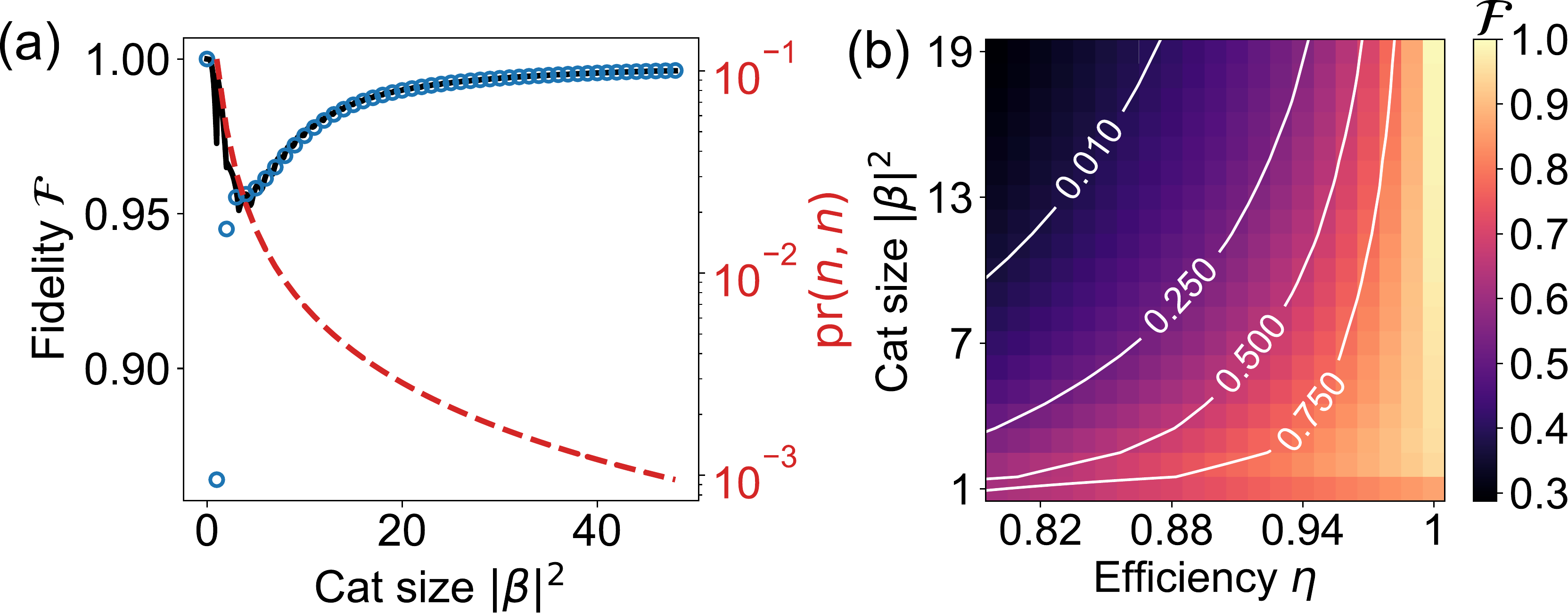}
	\caption{
	Two-component cat states. (a) The fidelity $\mathcal{F}$ of the cat state prepared by our scheme, $\ket{\psi_\mathrm{cat}}$, with respect to the ideal state $\ket{\mathrm{cat}_\beta}$. The blue circles are obtained from Eq.~\eqref{eqn:fid_cat} for $|\beta|^2 \in \mathbb{N}$. The continuous line is obtained from a numerical optimization of the parameters $\alpha$ and $n$ used in our scheme. The red dashed line shows the probability $\mathrm{pr}(n,n)$ of successfully heralding $\ket{\psi_{\mathrm{cat}}}$ using the optimal squeezing parameter.
	(b) The fidelity obtained with imperfect detection efficiency $\eta$ for various cat sizes when $\lambda = 0.82$ (10 dB of squeezing).
	White contour lines indicate the normalized volume of negativity in the prepared cat state's Wigner function.
	}
	\label{fig:catFid}

\end{figure}

Suppose we wish to prepare the even two-component Schr\"{o}dinger cat state $\ket{\mathrm{cat}_\beta}= \mathcal N \left( \ket{\beta} + \ket{-\beta} \right),$
where $\mathcal N=\left[2(1+e^{-2|\beta|^2})\right]^{-1/2}$ is the normalization factor.
In Sec.~\ref{sec:even_parity_det} we explained that the detected state $\ket{\chi}$ is a symmetrized version of the ancilla when the latter is in a coherent state.
Hence, one might think using $\ket{\phi} = \ket{\beta}$ would be the optimal choice to prepare $\ket{\mathrm{cat}_\beta}$.
However, while $\ket{\chi}$ would closely resemble $\ket{\mathrm{cat}_\beta}$, the remotely prepared state $\ket{\psi}$ would not have the desired photon-number distribution in the realistic scenario of finite squeezing.
As mentioned earlier, one can generally compensate for this effect by carefully choosing the control state coefficients. 
For cat states this compensation is experimentally easy since one can simply choose the coherent control state $\ket{\phi} = \ket{\alpha}$ with the amplitude $\alpha = \beta\lambda$. 
In this case, given the detection outcome $(n,n)$, we prepare the state
\begin{equation}
\begin{split}
    \ket{\psi_\mathrm{cat}} &= \mathcal{N}' \sum_{j=0}^{2n} \frac{\alpha^{2n-j}\lambda^{j}}{\sqrt{(2n-j)!}} A^*_{j,n}\ket{j} \\
    &=\mathcal{N}'\lambda^{2n} \sum_{j=0}^{2n} \frac{\beta^{2n-j}}{\sqrt{(2n-j)!}} A^*_{j,n}\ket{j},
\end{split}
    \label{eqn:heralded_cat}
\end{equation}
where $\mathcal{N}'$ is the new normalization factor.
To maximize the fidelity $\mathcal{F} = \left|\braket{\psi_\mathrm{cat}|\mathrm{cat}_\beta}\right|^2$, one should generally post-select on the detection outcome $n$ to be the closest integer to $|\beta|^2$.
This condition ensures that the photon-number distribution of $\ket{\psi_\mathrm{cat}}$ is centered on the flat portion of $A_{j,n}$.
In the particular case when the cat size $|\beta|^2$ is exactly an integer, i.e. $|\beta|^2=n$, the fidelity is given by
\begin{equation}
    \mathcal{F} = \frac{2^{2n+1}e^{-n}}{(1+e^{-2n})} \left ( \sum_{k=0}^{n} {n \choose k}^2  \frac{(2k)!}{n^{2k}} \right )^{-1}.
    \label{eqn:fid_cat}
\end{equation}
This fidelity asymptotically approaches unity with increasing cat size, as shown by blue circles in Fig.~\ref{fig:catFid}(a). For small $|\beta|^2$ (and hence small $n$), $A_{j,n}$ is not flat, causing the dip in the fidelity.

Further improvement of the fidelity for small $|\beta|^2$ can be obtained through numerical optimization\footnote{The optimization algorithms used Scipy's function \emph{optimize.minimize} with the default BFGS method.}
 of the parameters $\alpha$ and $n$. The optimized fidelity is shown by the black line in Fig.~\ref{fig:catFid}(a).
The small oscillations are due to discrete nature of the parameter $n$.

The effect of imperfect detectors on $\mathcal{F}$ can be numerically calculated using $\hat{\Pi}(\eta)$, as discussed in Sec.~\ref{sec:detEff}.
The result is shown in Fig.~\ref{fig:catFid}(b).
On the same plot, white contour lines indicate the volume of negativity, a well-known measure of nonclassicality that is obtained by integrating the negative regions of the cat state's Wigner function~\cite{kenfack2004negativity}.
We normalize the volume of negativity to that of an ideal cat state of equal size, i.e. 1 is the maximum amount of negativity for a cat state of that size, while 0 is no negativity.

Since our scheme requires post-selecting onto a single outcome $(n,n)$, it is important to consider the scaling of the probability of successfully preparing the cat state, which is given by $\mathrm{pr}(n,n) = \left\|\braket{\chi|\Psi}\right\|^2 = \left|\braket{\psi|\psi}\right|^2$. 
This probability depends on both $\beta$ and $\lambda$ since these parameters determine the number of photons after the beam splitter.
In fact, there is an optimal choice for $\lambda$ that maximizes $\mathrm{pr}(n,n)$ given a desired cat size $|\beta|^2$.
We numerically determine this optimal $\lambda$ by finding where the derivative of $\mathrm{pr}(n,n)$ with respect to $\lambda$ vanishes [red dashed line in Fig.~\ref{fig:catFid}(a)].
We find that $\mathrm{pr}(n,n)$ scales as $\sim|\beta|^{-5/2}$ which sets the fundamental limit on the success rate of our scheme.
Such decrease of the success rate with the cat size is a typical feature of post-selected schemes for large cat state preparation \cite{ourjoumtsev2007generation,etesse2015experimental,sychev2017enlargement}. Note the complementarity of our scheme with respect to Ref.~\cite{ourjoumtsev2007generation}: while the latter scheme requires a non-Gaussian photon-number input state and Gaussian homodyne measurement, our scheme requires a Gaussian input state and a non-Gaussian measurement.

\subsection{Four-component cat states}
\begin{figure}
    \centering
    \includegraphics[width=0.95\columnwidth]{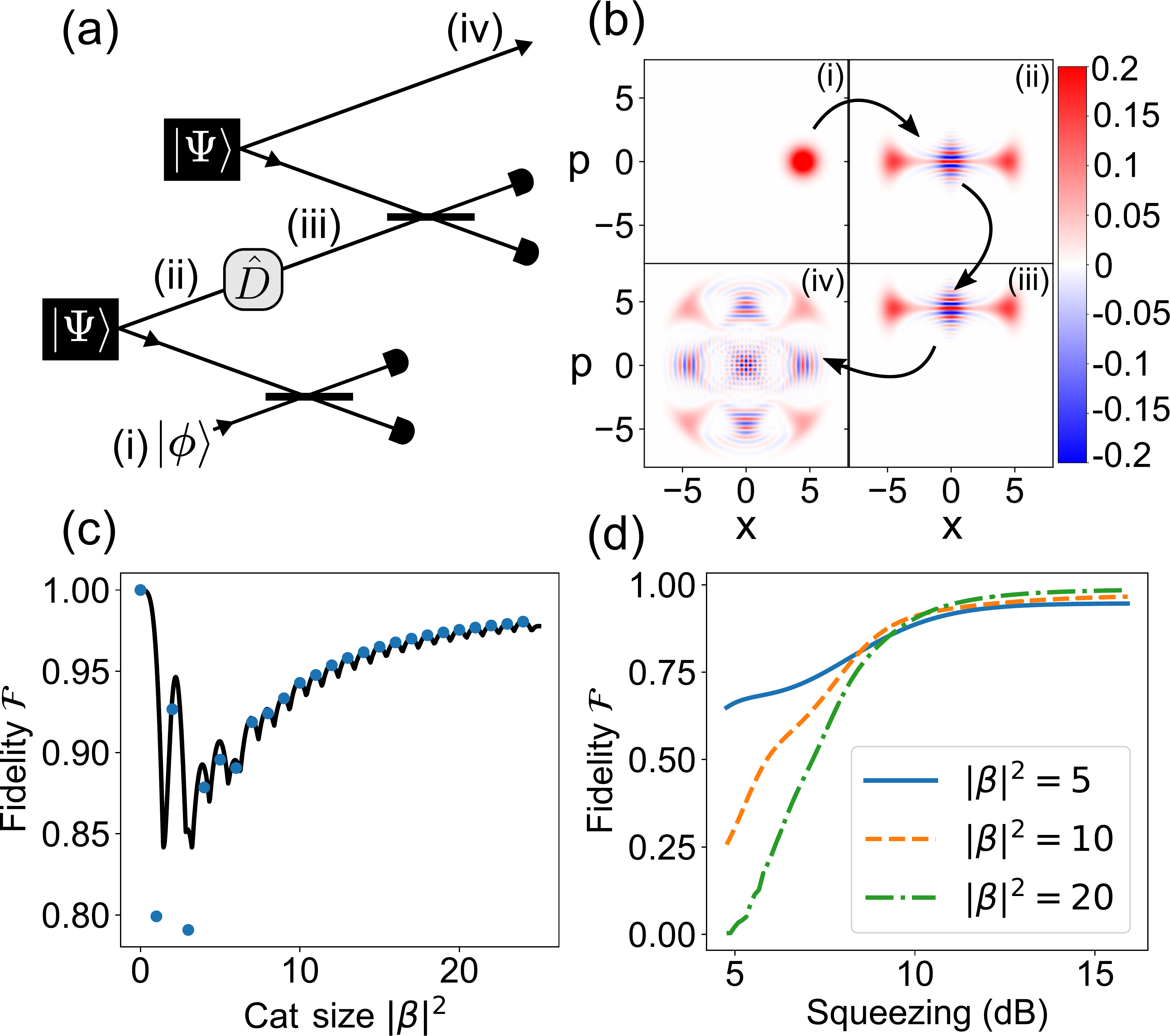}
	\caption{
	Four-component cat states. (a) We concatenate two instances of the even-parity detector. Between both instances, a displacement operation $\hat{D}$ is performed. Using this concatenated scheme, we prepare a four-component cat state.
	(b) The Wigner functions of the signal state is displayed for different stages of the procedure (here $\beta = \sqrt{10}$).
	(c) Fidelity of the four-component cat state with respect to the ideal state in Eq.~\eqref{eqn:catcodeState}, assuming infinite squeezing, i.e. $\lambda = 1$. 
	The blue circles are obtained by setting $n=|\beta|^2 \in \mathbb{N}$, whereas the line is obtained from a numerical optimization of the post-selected photon numbers at both stages, the amplitude of the initial coherent state and the displacement.
	(d) Behavior of the fidelity for finite squeezing, i.e. $\lambda < 1$.
	}
	\label{fig:catCode}
\end{figure}

It is possible to concatenate even-parity detectors in order to engineer more complex even-parity states, as shown in Fig.~\ref{fig:catCode}(a).
Here, we use such a concatenated scheme to prepare a superposition of four coherent states with equal amplitudes and equidistant phases, i.e. a four-component cat state.
These provide the logical states in the fault-tolerant quantum information ``cat code"~\cite{leghtas2013hardware,mirrahimi2014dynamically}.
Four-component cat states have also been studied in the context of quantum sensing~\cite{agarwal2004mesoscopic,dalvit2006quantum,roy2009subplanck} since they can have phase-space features that are smaller than Planck's constant~\cite{zurek2001sub}.
There exist schemes for preparing four-component cat states in the microwave domain with superconducting qubits~\cite{vlastakis2013deterministically,ofek2016extending} and in the motional state of a mechanical oscillator~\cite{howard2019quantum,ringbauer2018generation}.
To the best of our knowledge, there is no preparation scheme in the optical domain\footnote{While preparing this manuscript, we became aware of Ref.~\cite{hastrup2020deterministic} which proposes a deterministic scheme.}.

We begin by discussing our scheme assuming infinite squeezing, i.e. $\lambda = 1$.
The finite squeezing case will be considered later.
The first step is to produce a cat state $\ket{\beta} + \ket{-\beta}$ as described in the previous section. 
Next, we displace this cat state by $\beta$ in the direction perpendicular to the axis of the cat, i.e. apply the displacement operator $\hat{D}(i\beta)$:
\begin{equation}
\begin{split}
    \ket{\psi_{\mathrm{iii}}} &= \hat{D}(i\beta) \left( \ket{\beta} + \ket{-\beta}  \right) \\
    &= e^{i2|\beta|^2}\ket{\beta+i\beta} + e^{-i2|\beta|^2}\ket{-\beta+i\beta}.
\end{split}
\end{equation}
Finally, by measuring one of the modes of a fresh copy of $\ket{\Psi}$ with a second even-parity detector that uses $\ket{\psi_{\mathrm{iii}}}$ as its control state, we should prepare a symmetrized version of $\ket{\psi_{\mathrm{iii}}}$.
Indeed, post-selecting on the outcome $(2n,2n)$, we prepare an approximate version of the state
\begin{equation}
\begin{split}
    \ket{\psi_{\mathrm{iv}}} &= \ket{\beta-i\beta} + \ket{-\beta+i\beta} \\
    &+e^{-i2|\beta|^2}\left(\ket{\beta+i\beta} + \ket{-\beta-i\beta}\right),
\end{split}
\label{eqn:catcodeState}
\end{equation}
which is the desired four-component cat state.
It should be noted that the phase term $e^{-i2|\beta|^2}$ is determined by the size of the cat and cannot be independently controlled in our scheme.
We numerically simulated this procedure and plot the Wigner function of the state at each step in Fig.~\ref{fig:catCode}(b).

The fidelity of the final state with $\ket{\psi_{\mathrm{iv}}}$ as a function of its size is shown in Fig.~\ref{fig:catCode}(c). The behavior is similar to that to two-component cats studied above. That is, the fidelity asymptotically approaches unity with increasing $|\beta|^2$, but shows a dip for small $|\beta|^2$ due to variation of the Holland-Burnett state coefficients $A_{j,n}$.

We now consider the case of finite squeezing, i.e. $\lambda < 1$. 
If we were able to prepare a perfect two-component cat at stage (iii), the effect of finite squeezing on $\ket{\psi_{\mathrm{iv}}}$ could be compensated by appropriately choosing the amplitude and displacement of that cat.
However, unlike the first control state, this second control state is imperfect, which prevents the compensation from working properly for the following reason.
When $\lambda < 1$, the values of $j$ corresponding to the most significant coefficients of the heralded state's photon-number decomposition~\eqref{eqn:heralded_state} are shifted with respect to the center of the control state's photon-number decomposition, towards lower $j$. 
In other words, $\ket{\psi_{\mathrm{iv}}}$ is mainly determined by the photon-number coefficients $c_{2n-j}$ in the tail of the distribution of $\ket{\psi_{\mathrm{iii}}}$. 
While the fidelity of $\ket{\psi_{\mathrm{iii}}}$ is determined by its most significant photon-number coefficients, and increases with its size, the errors in the tail region of the distribution remain roughly constant.
As a result, the fidelity of the final state depends on the squeezing parameter, as shown in Fig.~\ref{fig:catCode}(d). That is,  high-fidelity compensation appears to be possible only above a squeezing level of about 12--13  dB, independent of the cat size. 

\subsection{Experimental realizations}

We briefly discuss our scheme in the context of current experimental capabilities. 
The requirements on the fidelity and size of the cat states ultimately depend on their desired use.
For instance, Fig.~\ref{fig:catFid}(b) suggests that it is possible to prepare large cat states exhibiting Wigner negativity using transition edge sensors, since these have $>$ 95\% efficiency and can detect up to $\sim$~20 photons~\cite{humphreys2015tomography}.

Besides the detectors, another important consideration is the two-mode squeezed vacuum source.
Preparing large cats in practice requires significant squeezing to obtain a reasonable heralding rate. 
At most 15 dB of squeezing is required to achieve the optimal heralding probability for two-component cat state of sizes up to $|\beta|^2=50$ [see red curve in Fig.~\ref{fig:catFid}(a)].
For example, at the 100 kHz experimental repetition rate usually used with transition edge sensors, one could herald a cat state of size $|\beta|^2=20$ at a rate of $\sim$~100 Hz using 13 dB of squeezing.
Such high squeezing levels are achievable in a pulsed experiment using optical waveguides~\cite{eto2011efficient,harder2016single} or in a continuous-wave experiment using an optical parametric oscillator~\cite{eberle2013stable,vahlbruch2016detection}.
If instead only e.g. 5 dB of squeezing is available, then one could herald a cat state of size $|\beta|^2=5$ at a rate of $\sim 1$ Hz.

Finally, we remark that imperfections such as thermal noise and modal purity of the two-mode squeezed vacuum source would also affect the performance of our scheme.
While the former imperfection can be modelled with optical loss~\cite{lvovsky2015squeezed} [see Fig.~\ref{fig:catFid}(b)], the latter is more complicated.
To minimize its effect, one would need to ensure that $\ket{\Psi}$ occupies a pair of well-defined spatio-temporal modes, one of which is well-matched to that of the control state $\ket{\phi}$.

\section{Summary and outlook}
To summarize, we devised an even-parity detector by exploiting the interference phenomenon that leads to the production of Holland-Burnett states in a time-reversed fashion.
The even-parity detector is controlled by varying the photon-number distribution of an ancillary control state. 
When this ancilla is in a coherent state, we showed that one can prepare two- and four-component Schr\"{o}dinger cat states of arbitrary size with nearly-perfect fidelity.
In practice, the size of the cats is limited by the dynamic range of photon-number resolving detectors and the linear optical losses.
Since these can detect up to $\sim 20$ photons with up to 95\% efficiency~\cite{lita2008counting}, we believe our scheme provides a promising route for preparing larger-scale cats in an experiment~\cite{thekkadath2019tuning}.

While this work focused on engineering superpositions of coherent states, our even-parity detector can prepare a wide range of even-parity states by using different types of control states.
For instance, it is straightforward to show that squeezed cats can be produced by using a control state in a squeezed coherent state, which, in turn, can be utilized to prepare Gottesman-Kitaev-Preskill states~\cite{vasconcelos2010all}.
Additionally, the use of photon subtraction or addition operations together with our scheme enables the preparation of odd-parity states.
Finally, our scheme to prepare four-component cat states exemplifies the ability to prepare states with more general discrete rotational symmetries in phase-space~\cite{grimsmo2019quantum,howard2019quantum} by concatenating even-parity detectors.

\acknowledgments{This work was supported by the Networked Quantum Information Technologies  Hub (NQIT) as part of the UK National Quantum Technologies Programme Grant EP/M013243/1. G.S.T. acknowledges funding from the Natural Sciences and Engineering Research Council of Canada (NSERC) and the Oxford Basil Reeve Graduate Scholarship.}

\appendix

\section{Imperfect detection efficiency}
\label{app:imp_det_eff}
Here, we derive the positive-operator valued measure element of the even-parity detector with imperfect detection efficiency. That is, we look for $\hat{\Pi}(\eta)$ such that 
\begin{equation}\label{prPOVM}
    \mathrm{pr}(n,n) = \mathrm{Tr}\left(\hat{\Pi}(\eta) \hat{\rho} \right).
\end{equation}
We model the imperfect detector efficiency by placing a fictitious beam splitter of transmissivity $\eta$ before each detector.
We then compute the probability of there to be $n$ photons in both transmitted modes when tracing over the reflected modes.
This probability can be obtained by performing a Bernoulli transformation on Eq.~\eqref{eqn:Pnn}~\cite{kiss1995compensation}:
\begin{equation}
\begin{split}
    \mathrm{pr}(n,n) &= \sum_{x=n}^{\infty}\sum_{y=n}^{\infty}{x \choose n}{y \choose n}\eta^{2n}(1-\eta)^{x-n}(1-\eta)^{y-n} \\
    &\times  \mathrm{Tr}\left( \ket{x,y}\bra{x,y}\hat{U}\hat{\rho}\otimes\ket{\phi}\bra{\phi}\hat{U}^\dagger \right). \\
\end{split}
\label{eqn:pnn_wLoss}
\end{equation}
Combining Eqs.~\eqref{prPOVM} and \eqref{eqn:pnn_wLoss} and using the cyclical property of the trace, we find
\begin{equation}
\begin{split}
    \hat{\Pi}(\eta) &= \sum_{x=n}^{\infty}\sum_{y=n}^{\infty}{x \choose n}{y \choose n}\eta^{2n}(1-\eta)^{x-n}(1-\eta)^{y-n} \\
    &\times \ket{\chi^{(x,y)}} \bra{\chi^{(x,y)}},
    \label{eqn:povm_imperfect_det}
\end{split}
\end{equation}
where $\ket{\chi^{(x,y)}} = \braket{\phi|\hat{U}^\dagger|x,y}$, which is a generalization of Eq.~\eqref{eqn:chi_full}.

\bibliographystyle{apsrev4-1}
\bibliography{refs}

\end{document}